\documentclass{PoS}

\usepackage[
   centertags, % (default) center tags vertically
   sumlimits,  %(default) Place the subscripts and superscripts of summation
   intlimits,  % Like sumlimits, but for integral symbols.
   namelimits, % (default) Like sumlimits, but for certain 'operator names' such as
]{amsmath} %
\usepackage{amssymb}

\makeatletter
\def\tagform@#1{\maketag@@@{\ignorespaces#1\unskip\@@italiccorr}}
\let\orgtheequation\theequation
\def\theequation{(\orgtheequation)}
\makeatother

\usepackage{dcolumn}
\usepackage{multirow}

\usepackage[english]{babel} 

\usepackage[numbers,square,sort&compress]{natbib}%

\usepackage[%
]{graphicx}

\newcommand{\beq}{\begin{equation}}
\newcommand{\eeq}{\end{equation}}

\newcommand{\gnuplotwidth}{0.8\textwidth}

\newcommand{\jpsi}{$J/\psi$ }
\newcommand{\unit}[1]{\, {\rm #1}}
\newcommand{\breakeq}{\nonumber\\}
\newcommand{\cc}{c \bar{c}}
\newcommand{\cpc}{c + \bar{c}}

\def\d{{\rm d}}

\title{Heavy quarks and charmonium at RHIC and LHC within a partonic transport model}

\ShortTitle{Heavy quarks and charmonium at RHIC and LHC}

\author{\speaker{Jan Uphoff}\\
        Institut f\"ur Theoretische Physik, Johann Wolfgang 
Goethe-Universit\"at Frankfurt, Germany\\
        E-mail: \email{uphoff@th.physik.uni-frankfurt.de}}

\author{Kai Zhou\\
Physics Department, Tsinghua University, Beijing, China}
       
\author{Oliver Fochler\\
       Institut f\"ur Theoretische Physik, Johann Wolfgang 
Goethe-Universit\"at Frankfurt, Germany}

\author{Zhe Xu\\
       Frankfurt Institute for Advanced Studies, Germany \\
Institut f\"ur Theoretische Physik, Johann Wolfgang 
Goethe-Universit\"at Frankfurt, Germany}

\author{Carsten Greiner\\
       Institut f\"ur Theoretische Physik, Johann Wolfgang 
Goethe-Universit\"at Frankfurt, Germany}

\abstract{
Heavy quark and charmonium production as well as their space-time evolution are studied in transport simulations of heavy-ion collisions at RHIC and LHC. In the partonic transport model \emph{Boltzmann Approach of MultiParton Scatterings} (BAMPS) heavy quarks can be produced in initial hard parton scatterings or during the evolution of the quark-gluon plasma. Subsequently, they interact with the medium via binary scatterings with a running coupling and a more precise Debye screening which is derived from hard thermal loop calculations, participate in the flow and lose energy. We present results of the elliptic flow and nuclear modification factor of heavy quarks and compare them to available data. Furthermore, preliminary results on $J/\psi$ suppression at forward and mid-rapidity are reported for central and non-central collisions at RHIC. For this, we study cold nuclear matter effects and the dissociation as well as regeneration of $J/\psi$ in the quark-gluon plasma.
}

\FullConference{XLIX International Winter Meeting on Nuclear Physics\\
		 24-28 January 2011\\
		 BORMIO, Italy}

\begin{document}

\section{Introduction}
Several experimental observations  indicate that a deconfined medium consisting of quarks and gluons -- the quark-gluon plasma (QGP) -- is produced in ultra-relativistic heavy-ion collisions \cite{Adams:2005dq,Adcox:2004mh}. Charm and bottom quarks are an ideal probe for the early stage of these collisions since they can only be created in initial hard parton scatterings of nucleon-nucleon interactions or in the medium in the beginning of the QGP phase, where the energy density is still large. After their production they interact with other particles of the medium and can, therefore, reveal important information about the properties of the QGP. Flavor conservation renders them as a unique probe since they are tagged by their flavor even after hadronization.

The experimental data of the elliptic flow $v_2$  and the nuclear modification factor $R_{AA}$  of heavy quarks \cite{Abelev:2006db,Adare:2006nq,Adare:2010de} show that the energy loss of charm and bottom quarks is comparable to that of light quarks. Whether this large energy loss is due to collisional or radiative interactions -- or both (or even other effects) -- is under investigation 
(see \cite{Adare:2010de} for a recent overview and comparison with data).

\jpsi suppression in heavy-ion collisions was proposed to be a signature of the QGP a long time ago \cite{Matsui:1986dk}, but it remains challenging to disentangle the contributions of hot and cold nuclear matter effects to the measured suppression (see e.g. Ref.~\cite{Brambilla:2010cs} for a recent overview). Since lattice calculation indicate that \jpsi can survive in the QGP to some extent \cite{Asakawa:2003re,Mocsy:2007jz}, a partonic transport model such as BAMPS is an ideal framework to investigate the influence of \jpsi dissociation and regeneration on the \jpsi suppression.

This article is organized as follows. After the introduction of the parton cascade BAMPS we will discuss the production of heavy quarks at RHIC and LHC. In Sec.~\ref{sec:energy_loss} our results on the elliptic flow  and nuclear modification factor of heavy quarks at RHIC are compared to the experimental data. Furthermore, preliminary results on \jpsi suppression are presented in Sec.~\ref{sec:jpsi} and in Sec.~\ref{sec:conclusions} we conclude with a short summary.

\section{Parton cascade BAMPS}
\label{sec:bamps}
For the simulation of the QGP phase we use the partonic transport model BAMPS \cite{Xu:2004mz,Xu:2007aa}, which stands for \emph{Boltzmann Approach of MultiParton Scatterings}.
BAMPS simulates the fully  $3+1$ space-time evolution of the QGP produced in heavy ion collisions by solving the Boltzmann equation,
\begin{equation}
\label{boltzmann}
\left ( \frac{\partial}{\partial t} + \frac{{\mathbf p}_i}{E_i}
\frac{\partial}{\partial {\mathbf r}} \right )\, 
f_i({\mathbf r}, {\mathbf p}_i, t) = {\cal C}_i^{2\rightarrow 2} + {\cal C}_i^{2\leftrightarrow 3}+ \ldots  \ ,
\end{equation}
dynamically for on-shell partons with a stochastic transport algorithm and pQCD interactions. ${\cal C}_i$ are the relevant collision integrals, and $f_i({\mathbf r}, {\mathbf p}_i, t)$ the one-particle distribution function of species $i=g,\, c,\, \bar{c},\, b,\, \bar{b},\, J/\psi$, since light quarks are not included yet. In addition to the binary collisions $2\rightarrow 2$, also $2\leftrightarrow 3$ scatterings for the gluons are possible. That is, the following processes are implemented in BAMPS:
\begin{align}
\label{bamps_processes}
	g+g &\rightarrow g+g \nonumber\\
	g+g &\rightarrow g+g+g \nonumber\\
	g+g+g &\rightarrow g+g	 \nonumber\\
	g+g &\rightarrow Q +\bar{Q} \nonumber\\
	Q+ \bar{Q} &\rightarrow g+g \nonumber\\
	g+Q &\rightarrow g+Q \nonumber\\
	g+\bar{Q} &\rightarrow g+\bar{Q} \nonumber\\
	J/\psi + g &\rightarrow \cpc  \nonumber\\
	\cpc &\rightarrow J/\psi + g
\end{align}
Details of the model and the employed cross sections can be found in \cite{Xu:2004mz,Xu:2007aa,Uphoff:2010sh}.

\section{Heavy quark production at RHIC and LHC}
In heavy-ion collisions charm and bottom quarks are produced in hard parton scatterings of primary nucleon-nucleon collisions or in the QGP. To estimate the initial heavy quark yield, we use PYTHIA \cite{Sjostrand:2006za} and scale from proton-proton collisions to heavy-ion collisions with the number of binary collisions. Secondary heavy quark production in the QGP is simulated with BAMPS. For the initial gluon distributions, the mini-jet model, the color glass condensate model and also PYTHIA are employed.

In Au+Au collisions at RHIC with $\sqrt{s_{NN}}=200 \, \rm{GeV}$ between 0.3 and 3.4 charm pairs are produced in the QGP, depending on the model of the initial gluon distribution, the charm mass and whether a $K=2$ factor for higher order corrections of the cross section is employed \cite{Uphoff:2010sh}. This is only a small fraction of the initially produced charm quarks and can be neglected for the most probable scenarios.

\begin{figure}
\centering
\includegraphics[width=\gnuplotwidth]{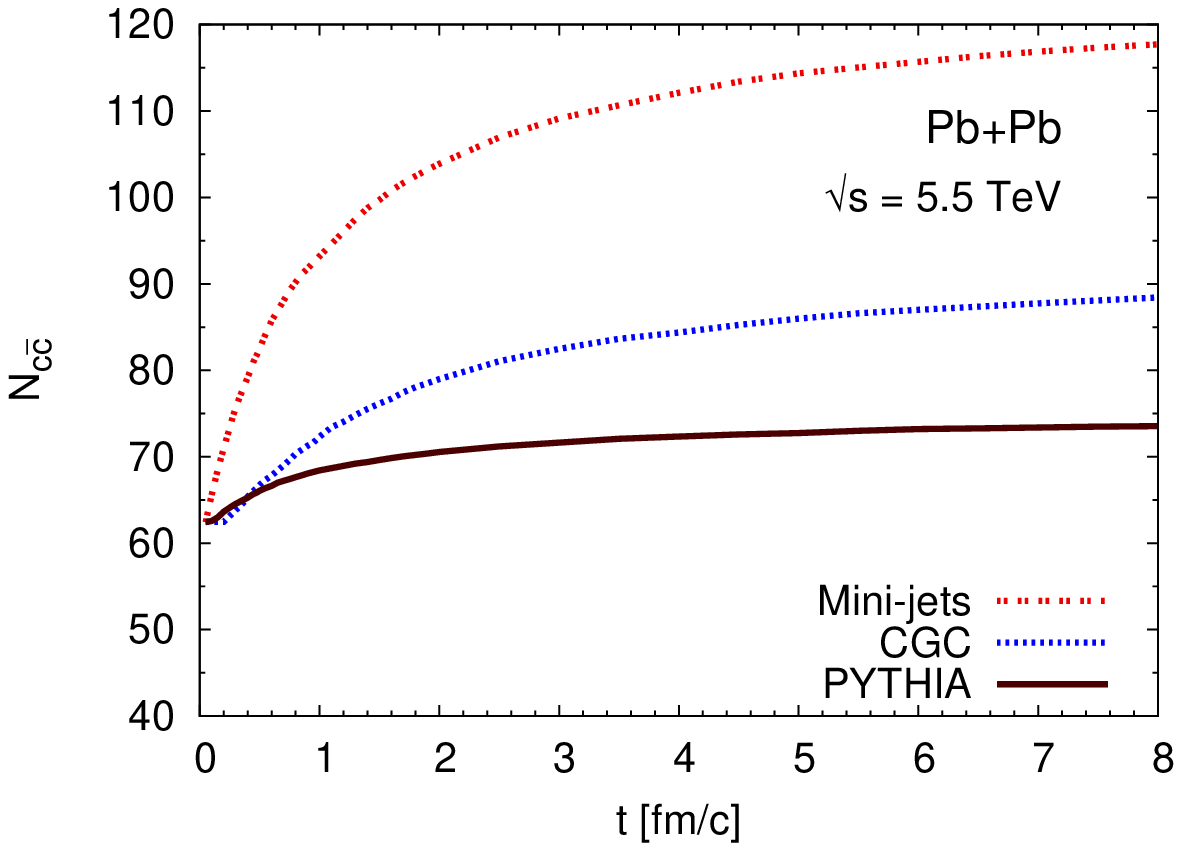}%eps

\includegraphics[width=\gnuplotwidth]{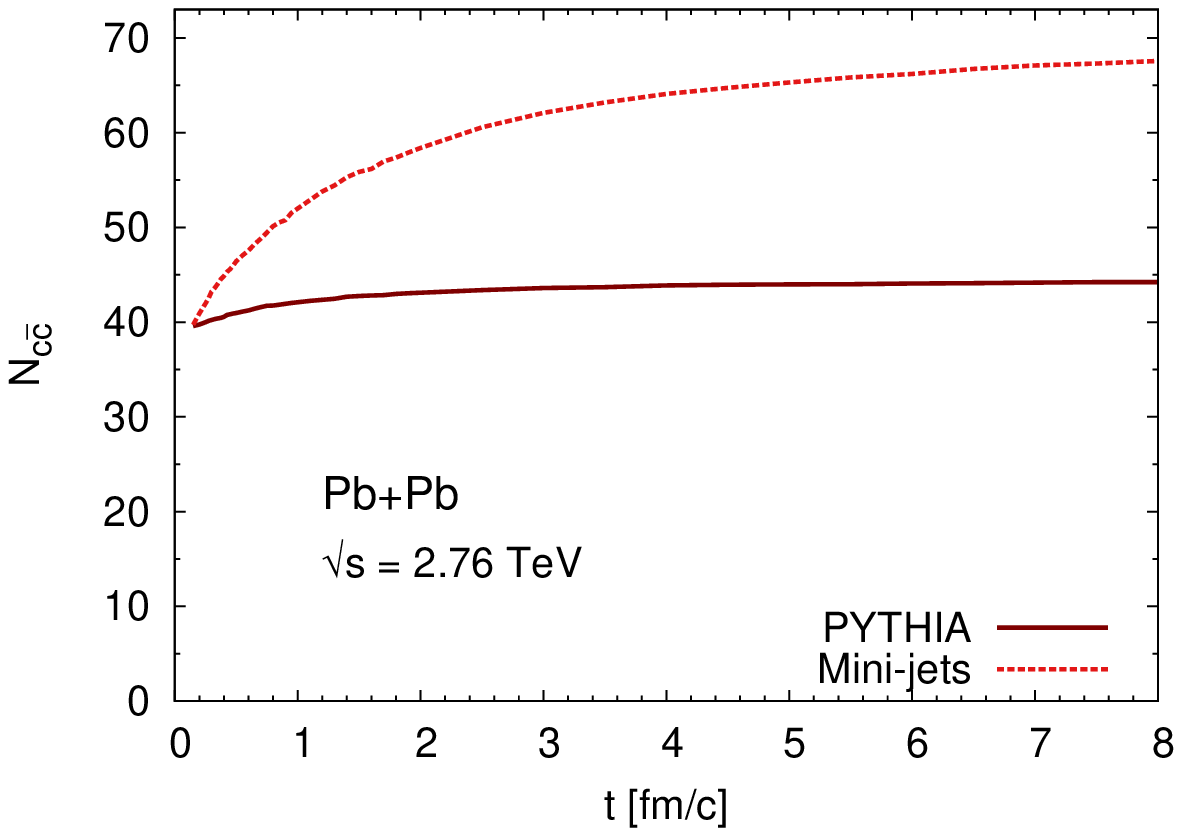}%eps
\caption{Number of charm quark pairs produced in a central Pb+Pb collision at LHC with $\sqrt{s_{NN}}=5.5 \, \rm{TeV}$ (top) and $\sqrt{s_{NN}}=2.76 \, \rm{TeV}$ (bottom) according to BAMPS. The initial parton distributions are obtained with PYTHIA and the mini-jet model (and the color glass condensate for $\sqrt{s_{NN}}=5.5 \, \rm{TeV}$). In all cases the initial charm quarks are sampled with PYTHIA for better comparison.}
\label{fig:charm_prod_276}
\end{figure}

At LHC with the much larger initial energy density, secondary charm production is enhanced and not negligible. In Pb+Pb collisions with $\sqrt{s_{NN}}=5.5 \, \rm{TeV}$ between 11 and 55 charm pairs are produced during the evolution of the QGP \cite{Uphoff:2010sh} (see left panel of Fig.~\ref{fig:charm_prod_276}). These values are of the same order as the initial yield. As is shown in the right panel of Fig.~\ref{fig:charm_prod_276}, even in the 2010 run with $\sqrt{s_{NN}}=2.76 \, \rm{TeV}$ between 5 and 28 charm pairs are created during the QGP phase.

Bottom production in the QGP, however, is very small both at RHIC and LHC and can be safely neglected. As a consequence, all bottom quarks at these colliders are produced in initial hard parton scatterings.

Further details on heavy quark production can be found in Ref.~\cite{Uphoff:2010sh}.

\section{Elliptic flow and nuclear modification factor of heavy quarks at RHIC}
\label{sec:energy_loss}

The elliptic flow and the nuclear modification factor
\begin{align}
\label{elliptic_flow}
  v_2=\left\langle  \frac{p_x^2 -p_y^2}{p_T^2}\right\rangle \ , \qquad \qquad
R_{AA}=\frac{{\rm d}^{2}N_{AA}/{\rm d}p_{T}{\rm d}y}{N_{\rm bin} \, {\rm d}^{2}N_{pp}/{\rm d}p_{T}{\rm d}y}
\end{align} 
($p_x$ and $p_y$ are the momenta in $x$ and $y$ direction in respect to the reaction plane)
of heavy quarks at mid-rapidity are observables which are experimentally measurable and reflect the coupling of heavy quarks to the medium. A large elliptic flow and a small $R_{AA}$ indicate strong interactions with the medium and a sizeable energy loss. Experimental results reveal that both quantities are of the same order as the respective values for light particles \cite{Abelev:2006db,Adare:2006nq,Adare:2010de}.

The leading order perturbative QCD cross section with a constant coupling $\alpha_s = 0.3$ and the regular Debye screening mass for  the $t$ channel is too small to build up an elliptic flow, which is in agreement with the experimental data \cite{Uphoff:2010fz}. However, if we take the running of the coupling into account and determine the screening mass from comparison to hard thermal loop calculations, we obtain an elliptic flow and $R_{AA}$, which are much closer to the data.

The following calculations are done analogously to \cite{Gossiaux:2008jv,Peshier:2008bg,Uphoff:2011ad}. An effective running coupling is obtained from measurements of $e^+e^-$ annihilation and non-strange hadronic decays of $\tau$ leptons \cite{Dokshitzer:1995qm,Gossiaux:2008jv}.
Since the $t$ channel of the $g Q \rightarrow g Q$ cross section is divergent, it is screened with a mass proportional to the Debye mass $m_{D}$:
\begin{align}
\label{t_screening}
   \frac{1}{t} \rightarrow \frac{1}{t-\kappa \, m_{D}^2}
\end{align}
The Debye mass is also calculated with the running coupling for consistency \cite{Uphoff:2011ad}. The prefactor $\kappa$ in Eq.~\ref{t_screening} is mostly set to 1 in the literature without a sophisticated reason. However, one can fix this factor  to $\kappa \approx 0.2$ by comparing the energy loss per unit length ${\rm d}E/{\rm d}x$ of the Born cross section with $\kappa$ to the energy loss within the hard thermal loop approach \cite{Gossiaux:2008jv,Peshier:2008bg,Uphoff:2011ad}.

\begin{figure}
\centering
\includegraphics[width=\gnuplotwidth]{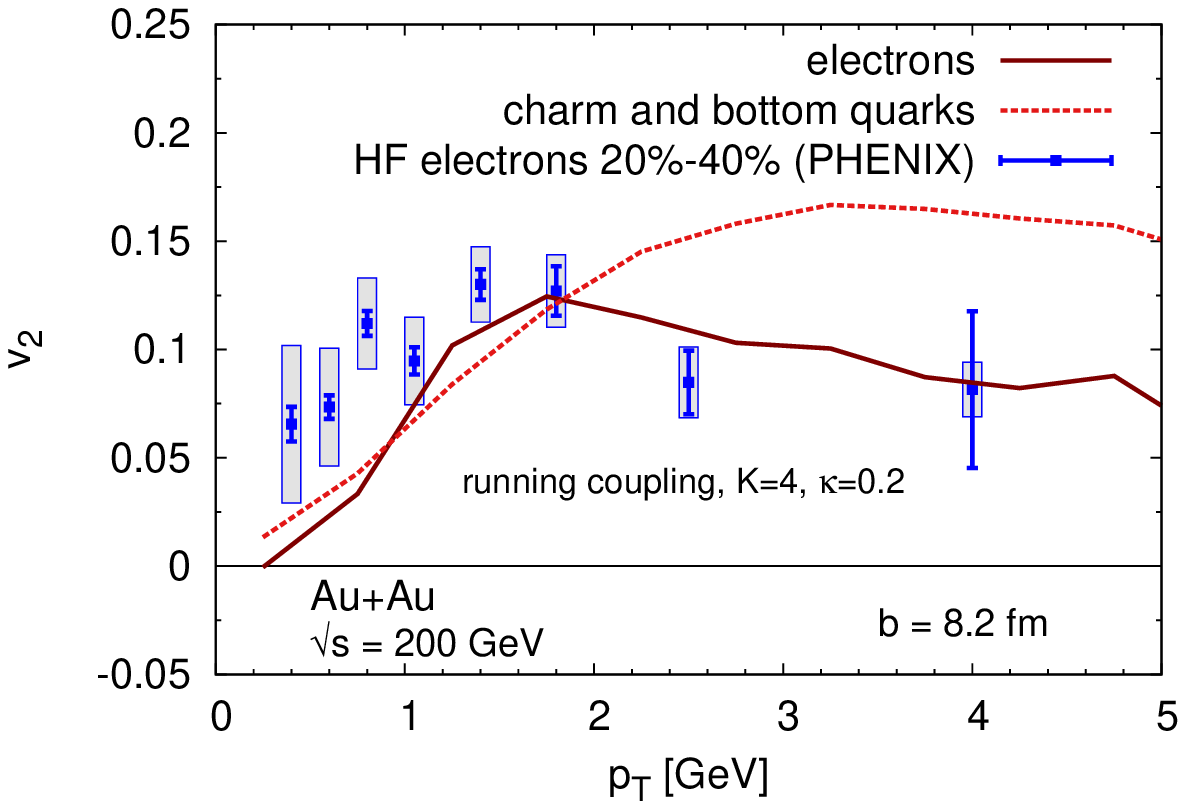}%eps

\includegraphics[width=\gnuplotwidth]{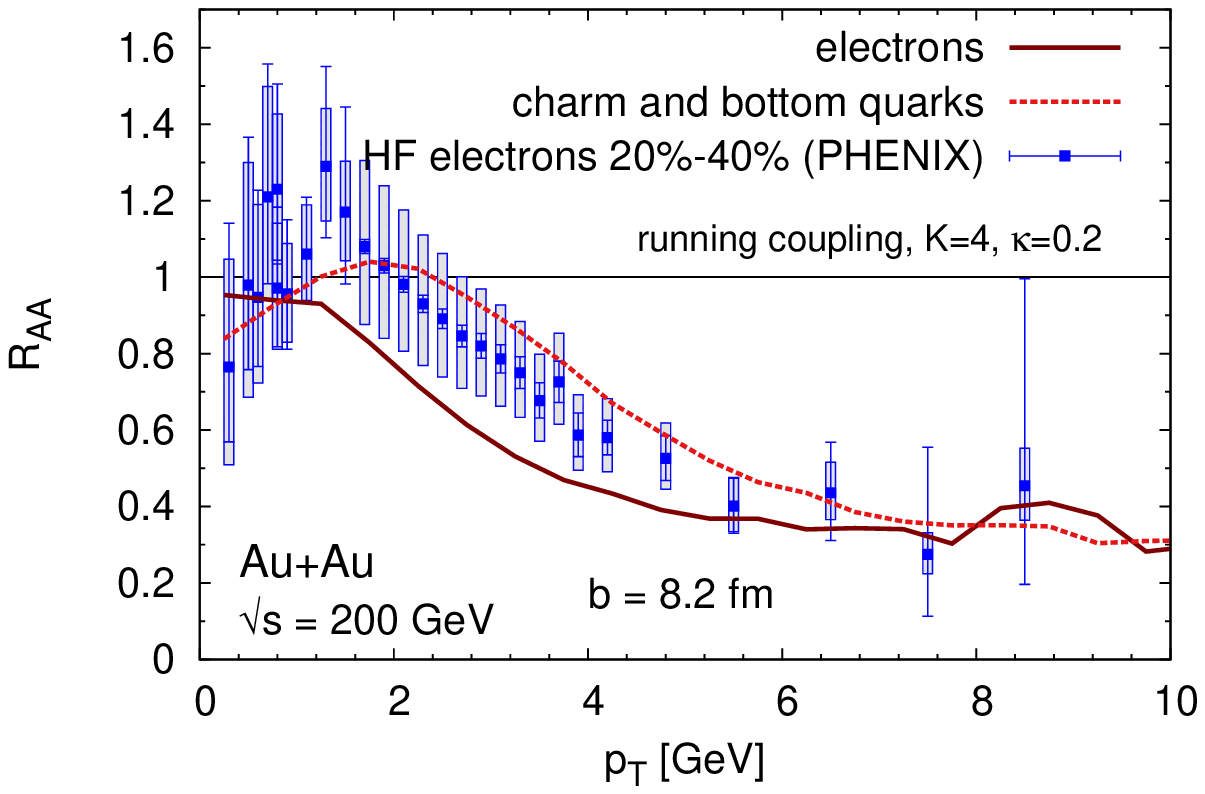}%eps
\caption{Elliptic flow $v_2$ (top) and nuclear modification factor $R_{AA}$ (bottom) of heavy quarks and heavy flavor electrons with pseudo-rapidity $|\eta|<0.35$ for Au+Au collisions at RHIC with an impact parameter of $b=8.2 \, {\rm fm}$. The cross section of $gQ \rightarrow gQ$ is multiplied with the factor $K=4$. For comparison, data of heavy flavor electrons \cite{Adare:2010de} is shown.}
\label{fig:v2_raa}
\end{figure}

This more accurate treatment increases the cross section of elastic gluon heavy quark scattering by about a factor of 10. Fig.~\ref{fig:v2_raa} shows the elliptic flow $v_2$ and nuclear modification factor $R_{AA}$ for heavy quarks and for heavy flavor electrons as a function of the transverse momentum $p_T$. 
To yield the same values for these variables as the experimental data, the leading order cross section of elastic collisions with the running coupling and improved Debye screening must still be multiplied by a $K$ factor of 4. We assume that this artificial $K$ factor stands for the contribution of radiative energy loss. However, it must be checked if these corrections have indeed a similar effect as a constant $K$ factor of 4. Therefore, the calculation of the next-to-leading order cross section is planned for the near future and will complement $2 \leftrightarrow 3$ interactions for gluons, which are already implemented in BAMPS \cite{Xu:2004mz}.

Especially for high $p_T$, the shape of the $v_2$ curve of heavy quarks is different from the experimental data. The reason for this discrepancy is that, experimentally, due to confinement, not heavy quarks, but heavy flavor electrons are measured. The latter stem from the decay of $D$ and $B$ mesons, which in turn are produced during hadronization of the QGP and which consist of a charm or bottom quark and a light quark. However, despite the hadronization and decay processes heavy flavor electrons still reveal information about heavy quarks. Essentially, the shape of their spectrum is the same as for heavy quarks, but shifted to lower $p_T$ due to the decay process. 

For the description of the hadronization process of charm (bottom) quarks to $D$ ($B$) mesons, we use  Peterson fragmentation \cite{Peterson:1982ak}. The decay to heavy flavor electrons is carried out with PYTHIA. The theoretical curves for heavy flavor electrons in Fig.~\ref{fig:v2_raa} are in good agreement with the experimental data for high $p_T$. For lower $p_T$ Peterson fragmentation is not a good description of the hadronization and another scheme like coalescence must be employed. In the coalescence picture, the light quarks of the $D$/$B$ mesons contribute also to its elliptic flow or nuclear modification factor, which increases both.

As a note, in contrast to previous results \cite{Uphoff:2010sy,Uphoff:2010bv} we employ here initial heavy quark distributions from MC@NLO \cite{Frixione:2002ik} and not from PYTHIA, since the former employs next-to-leading order processes for heavy quark production which describes p+p data more accurately \cite{Uphoff:2011ad}. This changes the $R_{AA}$ and $v_2$ slightly at intermediate $p_T$.

Studies on the $v_2$ and $R_{AA}$ of gluons in BAMPS are presented in \cite{Xu:2008av,Bouras:2008ip,Xu:2010cq,Fochler:2008ts,Fochler:2010wn}.

\section{$J/\psi$ suppression}
\label{sec:jpsi}

Two classes of phenomena are important for studying \jpsi suppression in heavy-ion collisions: cold nuclear matter effects and hot nuclear matter effects. The latter are effects that occur due to the presence of the quark-gluon plasma which is expected to be produced in ultra-relativistic heavy-ion collisions. Already 30 years ago \jpsi suppression due to melting in a medium was predicted to be a signature of the QGP \cite{Matsui:1986dk}. More recently, the regeneration of \jpsi from a charm and anti-charm pair received also much attention and is expected to be important at LHC and maybe also RHIC energies \cite{Grandchamp:2002wp,vanHees:2004gq,Thews:2005vj,Andronic:2006ky,Young:2009tj,Liu:2009nb,Zhou:2009vz}. 

Cold nuclear matter effects on the other hand are all phenomena of \jpsi suppression that would also be present if no QGP was formed. The best systems to study these effects are nucleon-nucleus collisions since one can measure directly the impact of nuclear effects on \jpsi production in the absence of a QGP. In ultra-relativistic heavy-ion collisions, however, both effects are present and it is a challenge to disentangle them.

In other words, cold nuclear matter effects influence the initial \jpsi production while hot nuclear matter effects describe the modification of the \jpsi yield during the evolution of the medium.

As a note, the presented ingredients of the model and our first results are preliminary. We tried to incorporate all important effects in the most accurate way. However, much more work on the details has to be done and uncertainties of employed parameters need to be studied.

The dominant process of initial \jpsi production is gluon fusion, $g+g \rightarrow J/\psi + g$. Therefore, the differential cross section for $J/\psi$ production in proton-proton (p+p) collisions is given by
\begin{align}
\label{jpsi_prod_cs_pp}
\frac{\mathrm{d} \sigma^{J/\psi}_{\rm pp}}{\mathrm{d}p_T \mathrm{d}y_{J/\psi} \mathrm{d}y_g }  =
x_1 x_2 f_g(x_1,\mu_F) f_g(x_2,\mu_F) 
\frac{\mathrm{d} \sigma_{gg \rightarrow J/\psi g}}{\mathrm{d}t } \ .
\end{align}

In the present paper we will parametrize the measured \jpsi production cross section in p+p collisions and incorporate cold nuclear matter effects as outlined in the following to get the production cross section for A+A collisions.

The most important contributions to cold nuclear matter effects are shadowing, nuclear absorption and the Cronin effect.
Shadowing describes the phenomenon that the parton distribution functions of partons in a nucleus $f^A(x,\mu_F)$ are modified compared to parton distribution functions in a nucleon $f^{\rm nucleon}(x,\mu_F)$. The ratio of both for parton $i$,
\begin{align}
	R_i^A(x,\mu_F) = \frac{f_i^A(x,\mu_F)}{A \ f_i^{\rm nucleon}(x,\mu_F)} \ ,
\end{align}
can be obtained, for instance, from nucleon-nucleus collisions. In the present study we employ the shadowing parametrization EPS08 \cite{Eskola:2008ca} for $R_i^A(x,\mu_F)$ and set the factorization scale $\mu_F=\sqrt{p_T^2+M_{J/\psi}^2}$ to the transverse mass of the $J/\psi$. Furthermore, we use a shadowing function that also depends on the transverse position in the collision \cite{Klein:2003dj}
\begin{align}
	\mathcal R_i^A ( {\bf x}_{T}, x, \mu_F) = 1 + N_{A,\rho} \, \left[R_i^A(x,\mu_F) - 1\right] \frac{T_{A}({\bf x}_{T})}{T_{A}(0)} \ .
\end{align}
$T_{A}({\bf x}_{T})$ denotes the nuclear thickness function from the Glauber model and $N_{A,\rho} = A \ T_{A}(0) / T_{AB}(0)$ is a normalization constant.

A produced \jpsi can also be absorbed by the remains of the collided nuclei shortly after its production. The nuclear absorption can be effectively described by a survival probability for \jpsi passing through nuclear matter
\begin{align}
	S_{\rm abs} = e^{ - \sigma_{\rm abs} \left[ T_A({\bf x}_{T}, z_A, +\infty) + T_A({\bf x}_{T} - {\bf b}, -\infty, z_B) \right]  } \ .
\end{align}
$T_A({\bf x}_{T}, z_1, z_2)$ is the nuclear thickness function, but not integrated over the full $z$ range $(-\infty,+\infty)$ but over the interval $[z_1,z_2]$. The limits in the formula above are chosen in such a way that it represents the path length which the produced \jpsi travels through the passing remains of the nuclei. For the absorption cross section we employ a value of $\sigma_{\rm abs} = 2.8 \unit{mb}$ \cite{Adare:2007gn}. However, the exact value or whether nuclear absorption is still present at RHIC and LHC energies is uncertain.

The \jpsi production cross section in p+p collisions can be parametrized as \cite{Liu:2009wza,Adare:2006kf}
\begin{equation}
\label{ppjpsi}
\frac{\d\sigma^{J/\psi}_{\rm pp}}{p_T \d p_T \d y} = \frac{2(n-1)}{D(y)} \left(1+{\frac{p_T^2}{D(y)}}\right)^{-n} \, {\frac{\d\sigma^{J/\psi}_{\rm pp}}{\d y}} \ ,
\end{equation}
where $n=6$ and $D(y)={\langle
p_t^2\rangle_{\rm pp}(n-2)}(1-y^2/Y^2)$.  $Y=\text{arccosh}(\sqrt{s_{\rm pp}}/(2 \,m_{J/\psi}))$ is the the maximum rapidity and $\langle p_t^2\rangle_{\rm pp}=4.14 \unit{GeV^2}$ the averaged transverse momentum squared. For $\d\sigma^{J/\psi}_{\rm pp}/\d y$ we employ a double Gaussian distribution \cite{Adare:2006kf,Liu:2009wza}. 

To account for the Cronin effect which describes the $p_T$ broadening of the fusing gluons in the nuclei, we add an additional path length dependence to the mean $p_T$ in the parametrization of the p+p cross section above:
\begin{align}
	\langle p_t^2 \rangle = \langle p_t^2 \rangle_{\rm pp} + a_{gN} \, L
\end{align}
with $a_{gN} = 0.1 \unit{GeV^2/fm}$ \cite{Zhao:2007hh} and 
\begin{align}
	L = \frac{1}{n_0} \left[  T_A({\bf x}_{T},-\infty, z_A) + T_A({\bf x}_{T} - {\bf b}, z_B, +\infty) \right]
\end{align}
which is the path length of the two incoming gluons through the nuclear matter of the other nucleus. $n_0$ denotes the maximum nuclear density from the Woods-Saxon distribution.

With all these cold nuclear matter effects the differential \jpsi production cross section can be written as 
\begin{align}
\frac{\mathrm{d} N^{J/\psi}_{\rm AA}}{\mathrm{d}p_T \mathrm{d}y_{J/\psi} \mathrm{d}y_g  } 
&= 
\int {\rm d} {\bf x}_{T} \int {\rm d}z_A \int {\rm d}z_B  \ 
n_A({\bf x}_{T}, z_A) \, n_A({\bf x}_{T} - {\bf b}, z_B)  \
\mathcal R_g^A ( {\bf x}_{T}, x_1, \mu_F) \, \mathcal R_g^A ( {\bf x}_{T} - {\bf b}, x_2, \mu_F) \breakeq
& \qquad
e^{ - \sigma_{\rm abs} \left[ T_A({\bf x}_{T}, z_A, +\infty) + T_A({\bf x}_{T} - {\bf b}, -\infty, z_B) \right]  }  \
x_1 x_2 f_g(x_1,\mu_F) f_g(x_2,\mu_F) 
\frac{\mathrm{d} \sigma_{gg \rightarrow J/\psi g}}{\mathrm{d}t }
\breakeq
&= 
\int {\rm d} {\bf x}_{T} \int {\rm d}z_A \int {\rm d}z_B  \ 
n_A({\bf x}_{T}, z_A) \, n_A({\bf x}_{T} - {\bf b}, z_B) \
\mathcal R_g^A ( {\bf x}_{T}, x_1, \mu_F) \, \mathcal R_g^A ( {\bf x}_{T} - {\bf b}, x_2, \mu_F) \breakeq
& \qquad
e^{ - \sigma_{\rm abs} \left[ T_A({\bf x}_{T}, z_A, +\infty) + T_A({\bf x}_{T} - {\bf b}, -\infty, z_B) \right]  } \
\frac{\mathrm{d} \sigma^{J/\psi}_{\rm pp}}{\mathrm{d}p_T \mathrm{d}y_{J/\psi} \mathrm{d}y_g }  \ ,
\end{align}
where Eq.~\eqref{jpsi_prod_cs_pp} was used.
In the following we assume that the emitted gluon in the \jpsi production process is soft. As a consequence, $x_1$ and $x_2$ are independent of its rapidity $y_g$ and we can integrate it out:
\begin{align}
\label{jpsi_ini_dist_cnm}
\frac{\mathrm{d} N^{J/\psi}_{\rm AA}}{\mathrm{d}p_T \mathrm{d}y_{J/\psi}} 
&= 
\int {\rm d} {\bf x}_{T} \int {\rm d}z_A \int {\rm d}z_B  \ 
n_A({\bf x}_{T}, z_A) \, n_A({\bf x}_{T} - {\bf b}, z_B) \
\mathcal R_g^A ( {\bf x}_{T}, x_1, \mu_F) \, \mathcal R_g^A ( {\bf x}_{T} - {\bf b}, x_2, \mu_F) \breakeq
& \qquad
e^{ - \sigma_{\rm abs} \left[ T_A({\bf x}_{T}, z_A, +\infty) + T_A({\bf x}_{T} - {\bf b}, -\infty, z_B) \right]  } \
\frac{\mathrm{d} \sigma^{J/\psi}_{\rm pp}}{\mathrm{d}p_T \mathrm{d}y_{J/\psi} } 
\end{align}

If there are no cold nuclear matter effects ($R_g^A = 1$, $a_{gN}=0$ and $\sigma_{\rm abs} = 0 \unit{mb}$) this formula simplifies to the well known binary scaling from p+p to A+A collisions:
\begin{align}
\frac{\mathrm{d} N^{J/\psi}_{\rm AA}}{\mathrm{d}p_T \mathrm{d}y } 
= T_{\rm AA} ({\bf b})
\frac{\mathrm{d} \sigma^{J/\psi}_{\rm pp}}{\mathrm{d}p_T \mathrm{d}y } 
\end{align}

The \jpsi distribution obtained with cold nuclear matter effects according to Eq.~\eqref{jpsi_ini_dist_cnm} is used as an input for our partonic transport model BAMPS, which we employ to study hot nuclear matter effects. In the medium, \jpsi can dissociate via the process $J/\psi + g \rightarrow \cpc$ whose cross section has been calculated to \cite{Peskin:1979va,Bhanot:1979vb}
\begin{align}
\label{cs_jpsi_g_ccb}
	\sigma_{ J/\psi \, g \rightarrow \cc} (s) = \frac{2^{11} \pi}{27} \frac{1}{\sqrt{M_c^3 \epsilon_{J/\psi}}} \frac{\left(\frac{w}{\epsilon_{J/\psi}}-1\right)^{3/2}}{\left(\frac{w}{\epsilon_{J/\psi}}\right)^5} \ ,
\end{align}
where $w = P_{J/\psi}^\mu P_{g\,\mu} / M_{J/\psi} = (s-M_{J/\psi}^2) / 2M_{J/\psi}$ is the gluon energy in the rest frame of the \jpsi and $\epsilon_{J/\psi} = 2M_D -  M_{J/\psi}$ is the binding energy of the $J/\psi$. The back reaction of this process, $\cpc \rightarrow J/\psi + g$, that is, the regeneration of \jpsi via charm anti-charm annihilation, is also taken into account in BAMPS. The cross section can be obtained from Eq.~\eqref{cs_jpsi_g_ccb} via detailed balance
\begin{align}
	\sigma_{\cc \rightarrow J/\psi \, g} (s) = \frac{4}{3} \frac{(s-M_{J/\psi}^2)^2}{s (s-4M_c^2)} \, \sigma_{ J/\psi \, g \rightarrow \cc} (s) \ .
\end{align}

Lattice results \cite{Asakawa:2003re,Mocsy:2007jz}  indicate that a \jpsi can survive in the QGP up to the dissociation temperature $T_d$ and melts at higher temperatures. In this study we use $T_d = 2T_c$ with $T_c = 165 \unit{MeV}$ being the phase transition temperature. In BAMPS we implement the effect in the following way. If the temperature in a cell is larger than the dissociation temperature $T_d$, all the \jpsi in this cell decay to charm and anti-charm quarks. This is a rather crude and somehow artificial treatment of this phenomenon. However, it could be improved by considering, instead of Eq.~\eqref{cs_jpsi_g_ccb}, a more sophisticated cross section for \jpsi dissociation, which leads to such a melting above $T_d$ by itself without the need of an additional cutoff. A first attempt in this direction has been done, for instance, in Ref.~\cite{Zhao:2007hh} with quasi-free scattering.

\begin{figure}
\centering
\includegraphics[width=\gnuplotwidth]{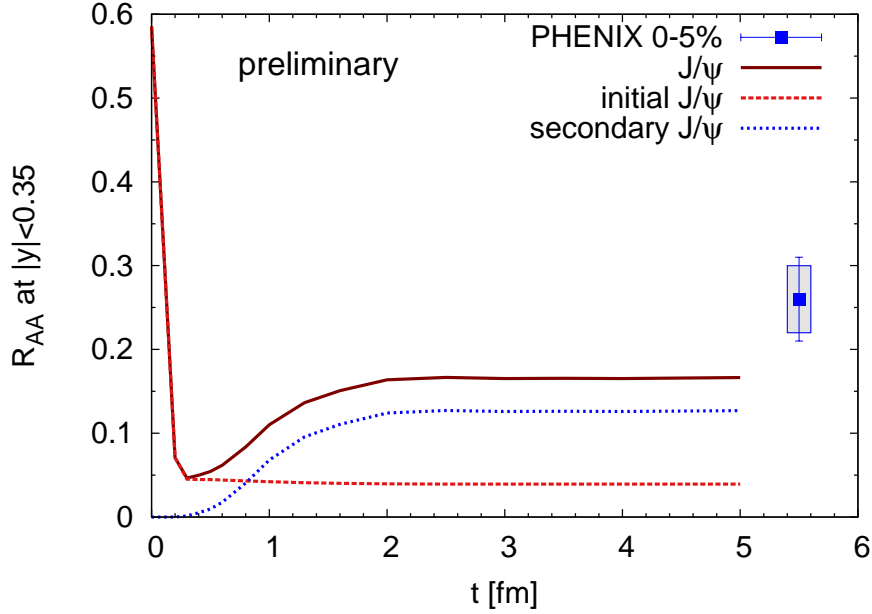}%eps
\caption{ $R_{AA}$ of \jpsi at mid-rapidity $|y| < 0.35$ in central Au+Au collisions at RHIC as a function of time. For comparison experimental data \cite{Adare:2006ns} is also shown.}
\label{fig:jpsi_raa_rhic_central}
\end{figure}

Figure~\ref{fig:jpsi_raa_rhic_central} shows the nuclear modification factor $R_{AA}$ of \jpsi for central Au+Au collisions at RHIC as a function of time.
The initial value is already smaller than 1 which is a consequence of cold nuclear matter effects. Then right from the beginning, a lot of initially produced charm quarks melt due to the large temperature of the QGP at RHIC. A counter-effect is the \jpsi regeneration which enhances the total \jpsi number. Our final value is a bit smaller than the experimental data point. However, there are significant uncertainties in the initial suppression. In addition, a different, more sophisticated  regeneration cross section could increase the regeneration and lead to a better agreement with the data point.

\begin{figure}
\centering
\includegraphics[width=\gnuplotwidth]{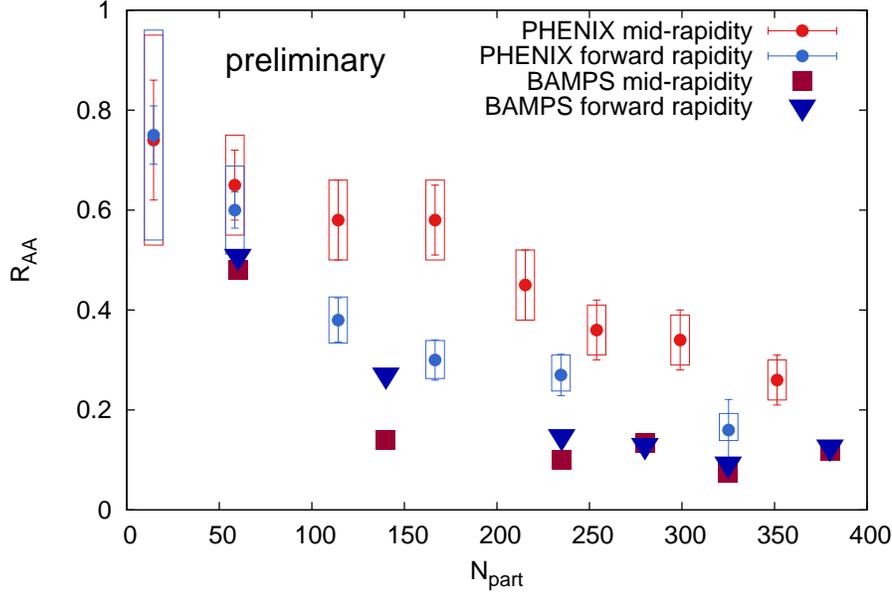}%eps
\caption{ $R_{AA}$ of \jpsi at mid-rapidity $|y| < 0.35$ for Au+Au collisions at RHIC as a function of the number of participants, together with experimental data \cite{Adare:2006ns}.}
\label{fig:jpsi_raa_rhic_npart}
\end{figure}

In Fig.~\ref{fig:jpsi_raa_rhic_npart} the $R_{AA}$ of \jpsi at RHIC is depicted as a function of the number of participants.
As we saw in Fig.~\ref{fig:jpsi_raa_rhic_central} for central collisions, our final $R_{AA}$ values lie also  for non-central collisions slightly below the experimental data points, but reproduce the overall shape of the data. This systematic underestimation indicates that either our suppression due to cold nuclear matter effects is too strong or the regeneration cross section is too small.

\section{Conclusions}
\label{sec:conclusions}
The production and evolution of heavy flavor particles have been studied within the partonic transport model BAMPS.
Charm and bottom production in the medium produced at RHIC are negligible compared to the initial heavy quark yield. At the LHC, however, secondary charm production can reach values comparable to the initial yield while bottom production in the QGP can also be neglected here.

The leading order cross section for heavy quark scatterings with particles from the medium is too small to explain the experimentally measured elliptic flow and nuclear modification factor. However, a more precise Debye screening and the explicit running of the coupling enhances the cross section and yields results for heavy flavor electrons, which are much closer to the data, although a $K$ factor of 4 must be employed for a good agreement with the data. In the future we will study if this simple multiplication of the cross section can indeed account for higher order contributions.

Furthermore, we investigated \jpsi suppression at RHIC. To estimate the initial \jpsi distribution we parametrized the p+p production cross section and took for the scaling to A+A collisions cold nuclear matter effects such as shadowing, nuclear absorption and the Cronin effect into account. The space-time evolution of \jpsi was carried out with BAMPS, which also allows dissociation and regeneration of $J/\psi$. Preliminary results of the nuclear modification factor $R_{AA}$ of \jpsi obtained with BAMPS for forward and mid-rapidity are slightly smaller than the measured data, but resemble the overall shape. Reasons for the smaller yield could be an overestimation of suppression due to cold nuclear matter effects or the small regeneration cross section. In a future study we will investigate this further and also perform calculations for the LHC.

\section*{Acknowledgements}
J.U. would like to thank A. Peshier for stimulating and helpful discussions and the kind hospitality at the University of Cape Town, where part of this work has been done.

The BAMPS simulations were performed at the Center for Scientific Computing of the Goethe University Frankfurt. This work was supported by the Helmholtz International Center for FAIR within the framework of the LOEWE program launched by the State of Hesse.

\bibliography{hq}

\end{document}